\documentstyle[multicol,epsf,prb,aps]{revtex}
\def\ruleleft{\vspace{-2.5\baselineskip}\begin{multicols}{2}\ \linebreak\vspace{-\baselineskip}\hrulefill\raisebox
{0.84mm}{$\!\rfloor$}\[\]\end{multicols}\vspace{-1.5\baselineskip}}
\def\ruleright{\vspace{-1.5\baselineskip}\begin{multicols}{2}\ \linebreak\raisebox
{-2.45mm}{$\lceil\!$}\hrulefill\end{multicols}\vspace{-\baselineskip}}
\newcommand{\rb}{\mbox{\boldmath$r$}}
\newcommand{\dd}{{\rm d}}
\begin{document}
\draft
\title{Density-functional theory for attraction between like-charged plates}
\author{Alexandre Diehl\\
{\small\it Departamento de F{\'\i}sica, Universidade Federal do Cear{\'a}, 
Caixa Postal 6030, CEP 60455-760, Fortaleza, CE, Brazil}\\
M. N. Tamashiro\\
{\small\it Materials Research Laboratory, University of California
at Santa Barbara, Santa Barbara, CA 93106-5130, USA}\\
Marcia C. Barbosa and Yan Levin\footnote{Corresponding author: 
levin@if.ufrgs.br}\\
{\small\it Instituto de F{\'\i}sica, Universidade Federal do Rio Grande do Sul,
Caixa Postal 15051, CEP 91501-970, Porto Alegre, RS, Brazil}\\}
\date{\today}
\maketitle
\begin{abstract}

We study the interactions between two negatively charged
macroscopic surfaces confining 
positive  counterions. A density-functional approach
is introduced which, besides the usual mean-field interactions, takes into
account the correlations in the positions of counterions.
The excess free energy  
is derived in the framework of the Debye-H{\"u}ckel 
theory of the one-component plasma, with the homogeneous density replaced by a 
weighted density. The minimization of the total free energy yields the density 
profile of the microions. The pressure is calculated and 
compared with the simulations 
and the results derived from integral equations theories. We find 
that the interaction between the two plates becomes attractive 
when their separation distance is sufficiently small and the surface charge 
density is larger than a threshold value.
\end{abstract}

\begin{multicols}{2}
\section{Introduction}

Solutions containing macromolecules are ubiquitous in the everyday life. From 
food colloids to the DNA, we are surrounded by these giant molecules which 
directly or indirectly govern every aspect of our lives.
In many cases the macromolecules in solution posses a net charge. The 
electrostatic repulsion between the
polyions is, often, essential to stabilization of colloidal 
suspensions. In the biological realm
the electrostatics is responsible for the condensation of the DNA and 
formation of actin bundles, while
various physiological mechanisms depend on the electrostatic interactions 
between the proteins and
the microions. In spite of their ubiquity, our understanding of 
polyelectrolyte solutions is far
from complete. 

The effort to fathom the role of electrostatics as it applies to the colloidal 
suspensions goes
back over half a century to the classic works of Derjaguin and 
Landau~\cite{DL41} and of Verwey and
Overbeek (DLVO).\cite{DLVO48} These in turn were based on the pioneering 
studies of
Gouy~\cite{Go10} and Chapman~\cite{Ch13} of double layers in metal electrodes. 
Following
these early contributions, a large effort has been devoted to solve the 
Poisson-Boltzmann (PB)
equation in various geometries. 
The mean-field treatment, based on the solution of the PB equation, suggests 
that the interaction between
two equally charged macroions in a suspension containing counterions is 
always repulsive.\cite{Is92,Sa94} In recent years, however, this dogma began 
to be questioned based
on simulations,\cite{Gu84} analytical
calculations~\cite{Kj84,Po89,Lo90,St90,Ro96,Pi98,Le99} and
experiments,\cite{Is86,Isr86,Cr96,Ca96} which indicated that for small 
distances and large charge
densities, two like-charged polyions might actually attract! 

The fundamental goal of this paper is to demonstrate that this attraction is 
linked to the correlations
between the microions omitted in the mean-field theories, and to establish the 
conditions under
which the attraction becomes possible. We shall consider the interaction 
between two infinite uniformly
charged plates confining their own point-like counterions. The mean-field 
approximation for this system
is obtained by solving the PB equation which, due to the planar symmetry, can 
be done analytically. Once
the density profile is obtained, all the other thermodynamic quantities can be 
easily derived. Thus, it
is not difficult to demonstrate that the pressure at the mean-field level, in 
units of energy,
is simply the density of counterions at the mid-plane between the plates. 
Since this is always positive,
no attraction is possible within the mean-field theory. 

Realization that the correlations between the counterions can strongly modify 
the mean-field
predictions goes back a number of years. One of the first approaches proposed 
by Kjellander and
Mar\v{c}elja~\cite{Kj84} was to include the correlations through the numerical 
solution of the
Anisotropic Hypernetted Chain Equation (AHNC). These authors found that the 
force per unit area
(pressure) can become negative in the presence of divalent counterions. Monte 
Carlo (MC) simulations
performed by Guldbrand {\it et al.}~\cite{Gu84} also indicate that as the 
surface charge density is
increased, the pressure decreases if the distance between the charged surfaces 
is sufficiently small. As
in the case of the AHNC calculations, attraction was found only in the 
presence of divalent
counterions. These authors, however, did not analyze the case of very 
high charge density and
short distance between the plates. In addition, since in the above 
calculations it is difficult to
separate the different physical contributions to the pressure, the mechanism 
that drives the
attraction remains unclear.

A different theoretical approach which attempted to shed some light on the 
mechanism of attraction
was advanced by Stevens and Robbins.\cite{St90} These authors proposed a 
density-functional theory
similar to the one often employed in studies of simple liquids. This approach 
introduces a
grand-potential free energy, $\Omega \left[\rho(\rb)\right]$, which is a 
functional of the non-uniform
density of counterions $\rho(\rb)$. The equilibrium properties of the system 
are obtained through
the minimization of the total free energy. The practical problem with this 
method is that the exact
form of the functional is not known. When the correlations between the 
microions are omitted,
the minimization of the grand potential, $\Omega_{\rm PB}$, becomes trivial 
and leads to the usual
PB equation.\cite{Sa94} In order to account for the correlations between the 
counterions, Stevens
and Robbins~\cite{St90} appealed to the Local Density 
Approximation (LDA).\cite{Ta85,Cu85,De89} Within
this approach an additional contribution, $f_{\rm LDA}$, is added to the 
mean-field expression,
$\Omega_{\rm PB}$. The expression for $f_{\rm LDA}$ adopted by Stevens and 
Robbins was obtained
through the extrapolation of the MC data for the {\it homogeneous}\/ 
One-Component Plasma
(OCP),\cite{Br79} but with the homogeneous density replaced by an {\it 
inhomogeneous}\/ density
profile. The minimization of the free-energy functional allowed them 
to determine the density
profile, $\rho(\rb)$, and the pressure, $P_{\rm OCP}$. The LDA, however, is 
not
without its own problems. The major drawback of this approach is that, for 
short distances and high charge
densities, the LDA is unstable. The reason for the instability is due to the 
fact that as
the density of counterions in the vicinity of the plates increases, the 
chemical potential decreases,
what attracts more particles to the region. This, in turn, leads to an 
unphysical ``chain reaction''
where all the counterions condense onto the plates. Clearly, when the distance 
between the counterions
becomes smaller than some threshold value, $s_{\rm corr}$, the LDA ceases to 
be a reliable
approximation.\cite{St90,Gr91,St96}

An improvement over the LDA is, the so called, Weighted Density Approximation 
(WDA).\cite{Ta85,Cu85,De89,Gr91}
In this case, the excess free energy is taken to be a function of an {\it 
average}\/ density,
$\rho_{w}(\rb)=\int \dd^3\rb'\,w(|\rb-\rb'|)\,\rho(\rb')$, averaged  
over a region of radius $s=s_{\rm corr}$, where the interactions between the 
counterions are
the strongest.\cite{Cu85,De89} The difficulty in the 
practical implementation of this scheme is the determination of a proper 
weight function. The simplest possible
form for $w(|\rb-\rb'|)$, used by Stevens and Robbins,\cite{St90,St96} was to 
assume that this function
has a long-range variation comparable to the wall separation.\cite{Rob} In 
this case, the
weighted density $\rho_w(\rb)$ is approximated by the homogeneous density 
independent of $\rb$. However,
when the walls are not close, $L>s_{\rm corr}$, the weighted function is no 
longer uniform and the
approximation adopted by Stevens and Robbins becomes unrealistic.
 
A beautiful explanation of the attraction between like-charged plates has been 
recently advanced by
Rouzina and Bloomfield.\cite{Ro96} These authors present a picture of 
attraction as arising from
the ground-state configuration of the counterions. Clearly at zero temperature 
the
counterions will
recondense onto the surface of the plates forming two intercalating Wigner 
crystals. The authors
advance a hypothesis that even at finite temperatures, relevant to the common 
experimental
conditions, the attraction is still governed by the zero-temperature 
correlations.
A somewhat different
formulations based on field-theoretic methodology have also been proposed. In 
these approximations
the attraction arises as a result of correlated fluctuations in the counterion 
charge
densities.\cite{Po89,Pi98,Le99} Although providing a nice qualitative 
explanation of the origin
of the attraction, these simple theories fail to yield a quantitative 
agreement
with the simulations.

In this paper we propose a different form of the weighted-density approach, 
which rectifies the
problems of the earlier theories while still remaining numerically tractable. 
The excess free
energy and the weight function, $w(|\rb-\rb'|)$, are derived from the 
Debye-H{\"u}ckel-Hole (DHH) theory of the
OCP.\cite{No84} The density profile is determined by minimizing the 
free-energy density
with respect to the {\it local}\/ density. Once the density profile is 
obtained,
the free energy of
the system is calculated by inserting it into the expression for the 
free-energy functional.
Given the free energy, all the thermodynamic properties of the system can be 
easily calculated. A
careful analysis of the behavior of the pressure as a function of the charge 
density and the distance
between the plates allows us to explore the nature and the origin of the 
attraction.

The remainder of the paper is organized as follows. The model and the PB 
approximation for the
density-functional approach are described in Sec.~\ref{poisson}. The WDA is 
introduced and
applied in Sec.~\ref{wda}. Our results and conclusions are summarized in 
Sec.~\ref{conclusion}.
 
\section{The Poisson-Boltzmann Approach}
\label{poisson}

\begin{figure}[ht]
\vspace*{2.5cm}
\begin{center}
\leavevmode
\epsfxsize=0.47\textwidth
\epsfbox[25 20 580 450]{"
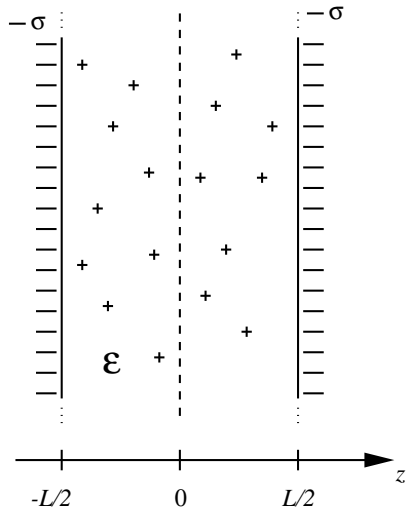"}
\end{center}
\vspace*{-2.5cm}
\begin{minipage}{0.48\textwidth}
\caption{Two infinite, negatively charged thin plates, with surface charge 
density
$-\sigma$ separated by distance $L$. The counterions are confined to the
region between the plates. The solvent is modeled as a
uniform medium of dielectric constant $\varepsilon$.}
\label{model}
\end{minipage}
\end{figure}

We consider two large, charged, thin surfaces each of area ${\cal A}$, 
separated by a 
distance $L$ (see Fig.~\ref{model}).
The two plates with a negative surface charge density, $-\sigma$, confine 
positive point-like monovalent counterions with charge $e$. 
The overall charge neutrality of the system is guaranteed by the 
constraint
\begin{equation}
\label{1}
\int_{-L/2}^{L/2}\dd z\,\rho(z)=\frac{2\sigma}{e},
\end{equation}
where $\rho(z)$ is the local number density of counterions and $z$ is the 
Cartesian coordinate
perpendicular to the plates. The space between the plates is assumed to be a 
dielectric continuum
of constant $\varepsilon$. 

In order to explore the thermodynamic properties of the system, we use a
density-functional approach. The grand potential of the system is
\begin{equation} 
\label{2}
\Omega \left[\rho\right] \equiv {\cal F}[\rho]-\mu  N\;,
\end{equation}
where $N$ is the total number of counterions, $\mu$ is their chemical 
potential and the functional
$\cal F$ is derived from the free-energy density of the homogeneous system, 
with the uniform density of counterions, $\rho_{\rm c}=N/L{\cal A}$, replaced 
by the local density $\rho(z)$.
For dilute systems, the ionic correlations can be neglected and the 
grand-potential functional (per unit
area) becomes
\end{multicols}
\ruleleft
\begin{equation}
\label{4}
\frac{\beta {\Omega}\left[\rho \right]}{{\cal A}}=
\int_{-L/2}^{L/2}\dd z\,\rho(z)\left\{\ln \left[\Lambda^3 
\rho(z)\right]-1\right\}
+\frac{\beta}{2} \int_{-L/2}^{L/2}\dd z\,\phi(z)\left[e\rho(z)+q(z)\right]
-\beta \mu \int_{-L/2}^{L/2}\dd z\, \rho(z)\;,
\end{equation}
\ruleright
\begin{multicols}{2}
where the electrostatic potential,
\begin{equation}
\label{4b}
\phi(\rb)=\int \dd^3\rb'\,\frac{e\rho (\rb')+q(\rb')}{\varepsilon |\rb-\rb'|},
\end{equation}
due to the symmetry of the problem, depends only on the $z$ coordinate. 
$\Lambda$ is the de Broglie thermal wavelength of the counterions,
$\beta=1/k_B T$ and $q(z)=-\sigma\left[\delta(z-L/2)+\delta(z+L/2)\right]$
is the surface charge density of the plates. The functional minimization of 
this
expression,
\begin{equation}
\label{5}
\frac{1}{\cal A}\frac{\delta\beta \Omega}{\delta\rho(z)}=0\;,
\end{equation}
produces the optimum density profile, 
\begin{equation}
\label{6}
\rho(z)={\rho}_0\,\exp \left[-\beta e\phi(z)\right]\;.
\end{equation}

The constant $\rho_0$ is determined from the overall charge-neutrality 
condition, Eq.~(\ref{1}),
\begin{equation}
\label{7}
{\rho}_0\equiv \frac{2\sigma}{e\displaystyle
\int_{-L/2}^{L/2}\dd z\,\exp\left[-\beta e\phi(z)\right]}\;.
\end{equation}

The electrostatic potential is obtained by solving the Poisson equation,
\begin{equation}
\label{8}
\frac{\dd^2\phi(z)}{\dd z^2}=-\frac{4\pi}{\varepsilon}\left[e\rho(z)+q(z)\right
],
\end{equation}
with the distribution of free ions given by Eq.~(\ref{6}). We find
\begin{equation}
\label{9}
\phi(z)=\frac{1}{\beta e}\ln \left[\cos^2 \left(\frac{z-z_0}{\lambda}
\right)\right]-\phi_0\;,
\end{equation}
where $\phi_0$ is the reference potential, which we will set to zero.
Here $\lambda=1/\sqrt{2\pi\lambda_B\rho_0}$ and 
$\lambda_B= \beta e^2/\varepsilon$ is the Bjerrum length.
Eq.~(\ref{8}) has to obey two boundary conditions, namely,
\begin{eqnarray}
\label{10}
E(z=0)&=&0\;,\nonumber\\
E\left(z=\pm \frac{L}{2}\right)&=&\pm\frac{4\pi\sigma}{\varepsilon}\;.
\end{eqnarray}
From the first equation, the electric field vanishes at the mid-plane
and, therefore, $z_0=0$. The second equation imposes the discontinuity of the 
electric field
at both charged surfaces, leading to
\begin{equation} 
\label{11}
\frac{1}{\lambda}\tan \left(\frac{L}{2\lambda}\right)=\frac{2\pi\sigma\lambda_B
}{e}\;.
\end{equation}
The potential at a point $z$ is, then, given by
\begin{equation}
\label{12}
\phi(z)=\frac{1}{\beta e}\ln \left[\cos^2 \left(\frac{z}{\lambda}\right)\right]
\;,
\end{equation}
with $\lambda$ the root of Eq.~(\ref{11}). The optimum density profile derived 
from this potential,
\begin{equation}
\label{13}
\rho(z)=\frac{{\rho}_0}{\cos^2(z/\lambda)}\;,
\end{equation}
can now be substituted into the free-energy functional, allowing the 
calculation of the total
free energy. The thermodynamic properties of the system can be determined from 
a suitable
differentiation of the total free energy. For example, the force between the 
two plates is given
by the minus derivative of the free energy with respect to the separation $L$ 
between the two surfaces.
This differentiation leads to a particularly simple expression for the force 
per unit of area
(or pressure),
\begin{equation}
\label{14}
\beta P={\rho}_0\;.
\end{equation}
We note that although it might be tempting to attribute this simple result to 
the contact theorem,
this is not the case, since the conditions under which this theorem holds are 
violated in the
present geometry; Eq.~(\ref{14}) is purely a mean-field result.

\section{The Weighted-Density Approximation}
\label{wda}

For dense systems, the correlations between the microions become relevant. For 
instance, if a
counterion is present at the position $\rb$, 
due to the electrostatic repulsion, the 
probability that
another counterion is located in its vicinity is drastically reduced. The 
correlations in the 
positions of the counterions reduce the mean-field estimate of the 
electrostatic
free energy.
No exact method exists for calculating this excess contribution. 
The
simplest approximation, the LDA, consists of adding to the Eq.~(\ref{4}) a 
{\it local}\/
functional, 
\begin{equation} 
\label{15}
f_{\rm LDA}=\int_{-L/2}^{L/2} \dd z\, \rho(z) f_{\rm 
corr}\left[\rho(z)\right]\;,
\end{equation}
where $f_{\rm corr}\left[\rho(z)\right]$ is the correlational free energy per 
particle. Within
the LDA one normally uses the expression derived for the homogeneous system, 
in which the
uniform density $\rho_{\rm c}=N/L{\cal A}$ is replaced by the local density 
profile $\rho(\rb)$.
Unfortunately, as was mentioned above, the LDA is unstable when the 
one-particle
density $\rho(\rb)$ is a rapidly varying function of the position. For 
example,
for high surface
charge densities, the minimization of the grand potential has no 
solution.\cite{Gr91}
To circumvent this and related problems intrinsic to the LDA, 
Tarazona~\cite{Ta85} and
Curtin and Ashcroft~\cite{Cu85} proposed a WDA, in which the free-energy 
density,
$f_{\rm LDA}$, is replaced by
\begin{equation} 
\label{16}
f_{\rm WDA}=\int_{-L/2}^{L/2} \dd z\, \rho(z) f_{\rm 
corr}\left[\rho_w(z)\right]\;.
\end{equation}
The fundamental difference between the LDA and the WDA is that the latter is 
assumed to
depend not on the local density $\rho(\rb)$, but on some average density 
within
the neighborhood of the point $\rb$,  
\begin{equation} 
\label{17} 
\rho_{w}(\rb)=\int\dd^3\rb' w\left[|\rb-\rb'| ; \rho(\rb)\right]\,
\rho(\rb')\; .
\end{equation}
This provides a control mechanism which prevents an unphysical, singular, 
buildup of concentration
at one point. The grand potential is obtained by adding the excess free energy 
per area,
given by Eq.~(\ref{16}), to the Eq.~(\ref{4}), 
\begin{eqnarray}
\label{18}
\frac{\beta\Omega \left[\rho\right]}{{\cal A}}&=&\int_{-L/2}^{L/2}\dd z\, 
\rho(z)\left\{\ln \left[\Lambda^3 \rho(z)\right]-1\right\}\nonumber\\
&&+\frac{\beta}{2}\int_{-L/2}^{L/2}\dd z\,\phi(z)\left[e\rho(z)+q(z)\right]\nonumber\\
&&+\beta\int_{-L/2}^{L/2} \dd z\, \rho(z) f_{\rm corr}
\left[\rho_w(z)\right]\nonumber\\ 
&&-\beta\mu\int_{-L/2}^{L/2}\dd z\,\rho(z)\;.  
\end{eqnarray}
Minimization of this expression leads to the optimum particle number density,
\begin{equation}
\label{19}
\rho(z)={\rho}_0 \exp \left[-\beta e\phi(z)-\beta \mu_{\rm 
ex}(z)\right]\;,
\end{equation}
where the excess chemical potential derived from $f_{\rm WDA}$, 
Eq.~(\ref{16}), is
\begin{eqnarray}
\label{20}
\mu_{\rm ex}(z)&=&\frac{\delta f_{\rm WDA}}{\delta \rho(z)}\nonumber\\
&=&f_{\rm corr}\left[\rho_w(z)\right]+ 
\int_{-L/2}^{L/2}\dd z'\,\rho(z')
\,\frac{\delta f_{\rm corr}\left[\rho_w(z')\right]}{\delta \rho(z)}\;,\nonumber\\
\end{eqnarray}
and the normalization coefficient is 
\begin{equation}
\label{20b}
\rho_0\equiv \frac{2\sigma}{e\displaystyle
\int_{-L/2}^{L/2}\dd z\,\exp\left[-\beta e\phi(z) -
\beta \mu_{\rm ex}(z)\right]} \; .
\end{equation}
The electrostatic potential satisfies the Poisson equation, Eq.~(\ref{8}), 
with the charge
density given by the Eq.~(\ref{19}). Integrating the Poisson equation over a 
rectangular shell
of area ${\cal A}$ and width $z$, and appealing to the Gauss' theorem, an 
integro-differential equation for
the electric field $E(z)$ can be obtained,  
\begin{equation}
\label{21}
{\cal E}({\bar z})=4\pi{\bar \sigma}\frac{\displaystyle \int_{0}^{\bar 
z}\dd{\bar z}'
\exp\left[-{\bar \mu}_{\rm ex}({\bar z}')+\int_{0}^{{\bar z}'}\dd{\bar 
z}''{\cal E}({\bar z}'')\right]}
{\displaystyle \int_{0}^{{\bar L}/2} \dd{\bar z}' \exp\left[-{\bar \mu}_{\rm 
ex}({\bar z}')
+\int_{0}^{{\bar z}'}\dd{\bar z}'' {\cal E}({\bar z}'')\right]}\;, 
\end{equation}
where  ${\cal E} \equiv e\beta\lambda_B E$, $\bar{\sigma}\equiv \sigma 
\lambda_B^2/e$,
${\bar z}\equiv z/\lambda_B$, ${\bar L}\equiv L/\lambda_B$ 
and $\bar{\mu}_{\rm ex}\equiv \beta\mu_{\rm ex}$.
The local density $\rho(z)$, which enters in 
the calculation of the excess chemical potential, Eq.~(\ref{20}), 
can be obtained from the derivative of the electric field, 
since $\nabla \cdot\mbox{\boldmath$E$}(\rb)=4 \pi 
e \rho(\rb)/\varepsilon$. The Eq.~(\ref{21}) explicitly 
fulfills the two
boundary conditions: ${\cal E}(0)=0$ and  ${\cal E}(\pm {\bar 
L}/2)=\pm 4\pi\bar{\sigma}$.  

The solution of this equation depends on the specific form of the excess 
free-energy density
and the weight function $w\left(|\rb-\rb'|\right)$. For the homogeneous OCP 
the electrostatic free energy can
be easily obtained using the DHH theory of Nordholm.\cite{No84} This is a 
simple linear
theory  based on the ideas of Debye and H{\"u}ckel. The electrostatic 
potential of the OCP is assumed
to satisfy a linearized PB equation. As a correction for the linearization, 
Nordholm postulated the existence
of an excluded-volume region of size $s_{\rm corr}$, from which all other ions 
are excluded.
The size of this region is such that the electrostatic repulsion between two 
counterions
is comparable to the thermal energy. Recent calculations using a generalized 
Debye-H\"uckel
theory indicate that this exclusion region is responsible for 
the oscillations observed
in the structure factor of the OCP at high couplings.\cite{Le299} 
Following Nordholm, we find
\begin{equation}
\label{24}
s_{\rm corr}=\frac{1}{\kappa_D}\left(1+3\lambda_B \kappa_D\right)^{1/3}-\frac{1
}{\kappa_D}\; ,
\end{equation}
where $\kappa_D=\sqrt{4\pi\lambda_B\rho_{\rm c}}$ is the inverse of the Debye 
length. The
excess free energy per particle is calculated to be
\end{multicols}
\ruleleft
\begin{equation}
\label{25}
\beta f_{\rm OCP}=\frac{1}{4}\left[1+\frac{2\pi}{3\sqrt{3}}
+\ln \left(\frac{\omega^2+\omega+1}{3}\right)-\omega^2 
-\frac{2}{\sqrt{3}}\tan^{-1}\left(\frac{2\omega+1}{\sqrt{3}}\right)\right]\;,
\end{equation}
\ruleright
\begin{multicols}{2}
where $\omega=(1+3\lambda_B\kappa_D)^{1/3}$. The correlational free energy 
per particle for the WDA, $f_{\rm corr}$,
which appears in (\ref{20}), is obtained by replacing $\rho_{\rm c}$ by 
$\rho_w(z)$
in the expression (\ref{25}), that is, 
$f_{\rm  corr}\left[\rho_w(z)\right]= f_{\rm OCP}\left[\rho_{\rm 
c}\to\rho_w(z)\right]$.

To obtain the weighted function~\cite{Ta85,Cu85} we require that the second 
functional
derivative of the free energy $\cal F$ in the limit of homogeneous densities,  
\begin{eqnarray}
\label{22}
\frac{\delta^2\beta{ \cal F}}{\delta \rho(\rb)\delta\rho(\rb')}
&=&\frac{\delta^3(\rb-\rb')}{\rho(\rb)}
+w(|\rb-\rb'|)\, \frac{\delta\beta\mu_{\rm ex}(\rb)}{\delta\rho(\rb')}\nonumber\\
&&+\frac{\lambda_B}{|\rb-\rb'|}\;,
\end{eqnarray}
produces the direct correlation function $C_2(\rb)$ of the homogeneous system, 
\begin{equation}
\label{23}
\frac{\delta^2\beta {\cal F}}{\delta \rho(\rb)\delta\rho(\rb')}
=\frac{\lambda_B}{|\rb-\rb'|}- C_2(\rb-\rb')\;.
\end{equation}
Following Groot,\cite{Gr91} we find that a reasonable approximation for the 
weight function is
\begin{equation} 
\label{26}
w(r)=w(|\rb|)=\frac{3}{2\pi s_{\rm corr}^2}\left(\frac{1}{r}-\frac{1}{s_{\rm 
corr}}\right)
\Theta(s_{\rm corr}-r)\;,
\end{equation}
where $\Theta(x)$ is the Heaviside step function. It is 
important to remember that the radius of the  
excluded-volume region, $s_{\rm corr}$, is now a function
of the position, since 
the average density $\rho_{\rm c}$, which appears in Eq.~(\ref{24}),  
is replaced  
by $\rho(z)$, the local density 
of counterions, see Eq.~(\ref{17}).
Taking advantage of the 
planar symmetry of the system,
the expression for the weighted density can be written explicitly as a 
one-dimensional quadrature,
\end{multicols}
\ruleleft
\begin{eqnarray}
\label{27}
\rho_w(z)&=&
\frac{3}{s_{\rm corr}^2}\int_{-L/2}^{L/2}\dd z'\;\rho(z')
\int_0^\infty \dd
\varrho\,\varrho\left(\frac{1}{\sqrt{\varrho^2+(z-z')^2}}-\frac{1}
{s_{\rm corr}}\right)\Theta\left(s_{\rm corr}-\sqrt{\varrho^2+(z-z')^2}\,\right)\nonumber\\
&=&\frac{3}{s_{\rm corr}^2}\int_{z_{<}}^{z_{>}}\dd z'\;\rho(z')
\int_0^{\sqrt{s_{\rm corr}^2-(z-z')^2}} \dd
\varrho\,\varrho\left(\frac{1}{\sqrt{\varrho^2+(z-z')^2}}-\frac1{s_{\rm corr}}
\right)\nonumber\\
&=&\frac{3}{2s_{\rm corr}^3}\int_{z_{<}}^{z_{>}}\dd z'\;\rho(|z'|) 
\left(s_{\rm corr}-|z-z'|\right)^2\;,
\end{eqnarray}
\ruleright
\begin{multicols}{2}
where $z_{<}\equiv \max(-L/2,z-s_{\rm corr})$, $z_{>}\equiv \min(L/2,z+s_{\rm 
corr})$
and $s_{\rm corr}$ is a function of $z$ through $\rho(z)$.

\section{Results and Conclusions}
\label{conclusion}

\begin{figure}[ht]
\begin{center}
\leavevmode
\epsfxsize=0.4\textwidth
\epsfbox[25 20 580 450]{"
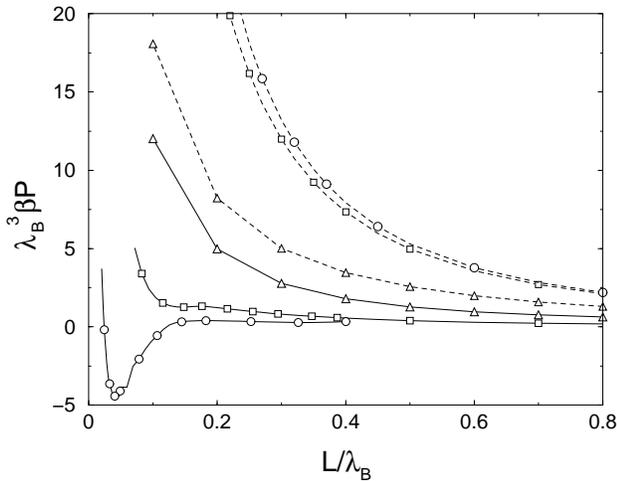"}
\end{center}
\vspace*{.7cm}
\begin{minipage}{0.48\textwidth}
\caption{The reduced osmotic pressure as a function of the plate separation 
for various surface
charge densities ${\bar \sigma}=\sigma \lambda_B^2/e$: 1 ($\triangle$), 5 
($\Box$) and 7 ($\bigcirc$).
The solid line is the WDA and the dashed line is the PB approximation for the 
same values
of $\bar \sigma$.} 
\label{wda_curvas}
\end{minipage} 
\end{figure}

Once $f_{\rm corr}$, $\mu_{\rm ex}(z)$ and $\rho_w(z)$ are defined, the 
electric field
and, consequently,
the optimum density profile can be determined from the numerical iteration of 
Eq.~(\ref{21}) until
convergence is obtained. The Helmholtz free energy, $F$, associated with 
the optimum
counterion distribution (\ref{19}),  is determined by substituting it into the
free-energy functional ${\cal F}$,
\end{multicols}
\ruleleft
\begin{equation}
\label{28}
\frac{\beta F}{{\cal A}}=\frac{2\sigma}{e} \left[\ln \left(\Lambda^3{\rho}_0\right)-1\right]
-\frac{\beta e}{2} \int_{-L/2}^{L/2} \dd z\,\rho(z) \phi(z)-
\beta\sigma\phi\left(\frac{L}{2}\right) 
-\int_{-L/2}^{L/2} \dd z\,\rho(z) \left\{\beta\mu_{\rm ex}(z) 
-\beta f_{\rm corr}\left[\rho_w(z)\right]\right\} \;.
\end{equation}
\ruleright
\begin{multicols}{2}
Using this expression, the pressure, for different distances between the 
plates, $L$, and
various charge densities, $\sigma$, can be easily obtained through numerical 
differentiation,
\begin{equation}
\label{29}
 P=-\frac{1}{\cal A}\frac{\partial {F}}{\partial L}\;,
\end{equation}
as shown in Fig.~\ref{wda_curvas}. When the charge density is below a 
threshold value,
$\bar{\sigma}<\bar{\sigma}_{c}$, the dimensionless pressure, $\lambda_B^3 
\beta P$, is always
positive and a monotonically decreasing function of $\bar{L}$. Above the 
critical surface charge
density the pressure exhibits a distinct minimum. In particular, we find that 
for sufficiently
high surface charge densities the force between the two like-charged surfaces 
becomes negative, i.e.
the two plates attract!

\begin{figure}[ht]
\begin{center}
\leavevmode
\epsfxsize=0.4\textwidth
\epsfbox[25 20 580 450]{"
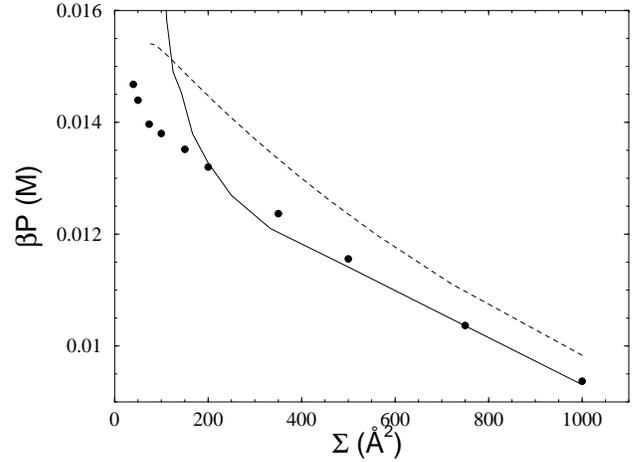"}
\end{center}
\vspace*{.4cm}
\begin{minipage}{0.48\textwidth}
\caption{Variation of $\beta P$ with $\Sigma=e/\sigma$ for $L=150$ {\AA}:
from  PB (dashed), WDA (solid)
and AHNC ($\bullet$) from Ref.~[8].} 
\label{stev} 
\end{minipage}
\end{figure}

In order to compare our results with other theories,\cite{Kj84,St90} we 
assumed that the dielectric
medium between the plates is water at room temperature and, consequently, that 
the
Bjerrum length is $\lambda_B=7.14\,${\AA}. The distance between the plates is 
fixed
at 150 {\AA} and the
inverse of the surface charge density, $\Sigma=e/\sigma$, is varied from 
40 {\AA}$^2$ to 1000 {\AA}$^2$.
Our results, illustrated in Fig.~\ref{stev}, show that for small surface 
charge densities the pressure
increases almost linearly with the inverse charge density, $\Sigma$. In this 
case, since
$P_{\rm corr} \ll P_{\rm PB}$, the pressure is dominated by the PB behavior. 
However, when
the charge density becomes large, the slope of $P_{\rm WDA}$ increases due to 
the strong repulsion
between the counterions.

\begin{figure}[ht]
\begin{center}
\leavevmode
\epsfxsize=0.4\textwidth
\epsfbox[25 20 580 450]{"
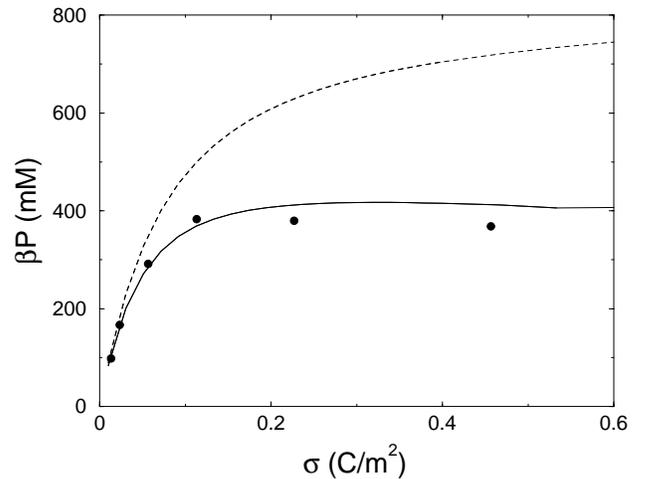"}
\end{center} 
\vspace*{.6cm}
\begin{minipage}{0.48\textwidth}
\caption{The osmotic pressure as a function of the surface charge density, 
when the
distance between the plates is fixed at  $L=21$ \AA, from PB (dashed) and  WDA 
(solid). The circles ($\bullet$) are the data from Ref.~[7].}
\label{guld}
\end{minipage}
\end{figure}

We also compare our calculations with the simulations of Guldbrand {\it et 
al.}\cite{Gu84} In this
case, the distance between the plates is fixed at 21 {\AA} and the surface 
charge density is
varied from 0.01 C/m$^2$ to 0.6 C/m$^2$, as shown in Fig.~\ref{guld}. 

\noindent When the density of counterions is small, $P_{\rm WDA}$ does not differ significantly from
$P_{\rm  PB}$. As the surface
charge density is increased, the correlations among the counterions become 
relevant and
$P_{\rm WDA}$ changes its slope and begins to decrease. Our 
results are in good agreement
with the simulations, which also indicate that for a separation of 21 {\AA} 
the
pressure exhibits a
region where it decreases with increase in the surface charge 
density.\cite{Gu84}

\acknowledgments

We acknowledge the fruitful discussions with Marcelo Louzada-Cassou, Rudi 
Podgornik, Roland Kjellander
and Stjepan Mar\v{c}elja. One of us, Marcia Barbosa, is particularly grateful 
for the useful
discussion with Mark O. Robbins. This work was supported in part by CNPq --- 
Conselho Nacional de
Desenvolvimento Cient{\'\i}fico e Tecnol{\'o}gico and FINEP --- Financiadora 
de Estudos e Projetos,
Brazil. This research was also supported by the National Science Foundation 
under Grant No. PHY94-07194.

\end{multicols}

\end{document}